\documentclass{aa}
\usepackage{graphics}
\usepackage{times}

\renewcommand{\ion}[2]{\mbox{#1\ {\scriptsize #2}}}

\def\logg{$\log g$\thinspace}

\def\hh{H1504+65}

\newcommand{\Teff}{\hbox{$T_{\rm eff}$}}

\newcommand{\lppr}{\stackrel{<}{\scriptstyle \sim}}
\newcommand{\lappr}{\raisebox{-0.4ex}{$\lppr $}}

\def\etal{{et\thinspace al.}\ }

\def\deg{\ifmmode ^\circ \else $^\circ$ \fi }
\def\solar{\ifmmode _{\mathord\odot}\else $_{\mathord\odot}$\fi}
\begin{document}

   \thesaurus{07 
    ( 08.01.1; 08.01.3; 08.05.3; 08.16.4; 08.23.1, 08.09.2 H1504+65)}

   \title
   {The EUV spectrum of the unique bare stellar core H1504+65\thanks
   {Based on observations with the Extreme Ultraviolet Explorer (EUVE) and
ROSAT. Some of the data presented herein were obtained at the W.M.\,Keck
Observatory, which is operated by the California
Institute of Technology and the University of California.}}
 
  \subtitle{}

   \author{Klaus Werner\inst{1} \and Burkhard Wolff\inst{2}}
   \offprints{K. Werner}
   \mail{werner@astro.uni-tuebingen.de}
   \institute{Institut f\"ur Astronomie und Astrophysik, 
              Universit\"at T\"ubingen, 
              D-72076 T\"ubingen, Germany 
              \and
              Institut f\"ur Theoretische Physik und Astrophysik,
              Universit\"at Kiel,
              D-24098 Kiel, Germany
             }

   \date{Received date; accepted date}
 
   \maketitle

   \begin{abstract}
We performed a spectral analysis of the EUVE spectrum and ROSAT data of the
unique object \hh, which is an extremely hot post-AGB star entering the white
dwarf cooling sequence. It is the only pre-white dwarf known, whose surface is
free of hydrogen and helium, hence, it represents the bare core of a former AGB
star. The EUV spectrum (75--150\AA) is dominated by strong \ion{O}{VI} lines
and we can identify a number of \ion{Ne}{VII} lines.

EUVE and ROSAT data can be fitted with models between \Teff=170\,000\,K and
200\,000\,K. We derive an extraordinarily high neon abundance (2\%--5\% by
mass, i.e.\ 20--50 times solar) which we confirm by an optical \ion{Ne}{VII}
line detected in a Keck HIRES echelle spectrum. This abundance is expected for
$3\alpha$ processed matter and corroborates our understanding of \hh\ as a C-O
stellar core which has lost its entire H- and He-rich envelopes.

\keywords{ 
           Stars: abundances -- 
           Stars: atmospheres  -- 
           Stars: evolution -- 
           Stars: AGB and post--AGB --
           white dwarfs --
           Stars: individual: H1504+65
         }
   \end{abstract}

\section{Introduction} 

The optical counterpart of the soft X-ray source \hh\ was detected by Nousek
\etal (1986) and spectroscopy revealed the unique nature of this object. From
the spectroscopic signatures it belongs to the PG\,1159 stars, which are very
hot (\Teff=75\,000--180\,000\,K) hydrogen-deficient pre-white dwarfs
(\logg=5.5--8, cgs-units). But in contrast to all other PG\,1159 stars it
appeared to be also helium-deficient, which was later confirmed by model
atmosphere analyses of optical spectra (Werner 1991) and FUV spectroscopy with
the {\it Hopkins Ultraviolet Telescope} (Kruk \& Werner 1998). Detailed NLTE
line profile fitting revealed that \hh\ is the most massive PG\,1159 star,
having the highest surface gravity, and that it is among the hottest of this
group (M/M$_{\sun}=0.86\pm 0.15$, \logg=$8.0\pm 0.5$, \Teff=170\,000\,K$\pm$
20\,000\,K; Werner 1991).

\hh\ is one of the brightest sources in the EUV and an early attempt to
understand its spectrum recorded with the {\it Extreme Ultraviolet Explorer}
(EUVE) failed in several respects, particularly the model flux was
overestimated by an order of magnitude, however,  it became obvious that the
spectral appearance is dominated by strong lines of \ion{O}{VI} (Barstow \etal
1995). Since that time considerable progress was made in NLTE modeling of
stellar atmospheres so that a new attempt to analyze these data together with
soft X-ray ROSAT observations and new optical Keck spectra seemed promising.

\begin{figure*}
  \resizebox{0.965\hsize}{!}{\includegraphics{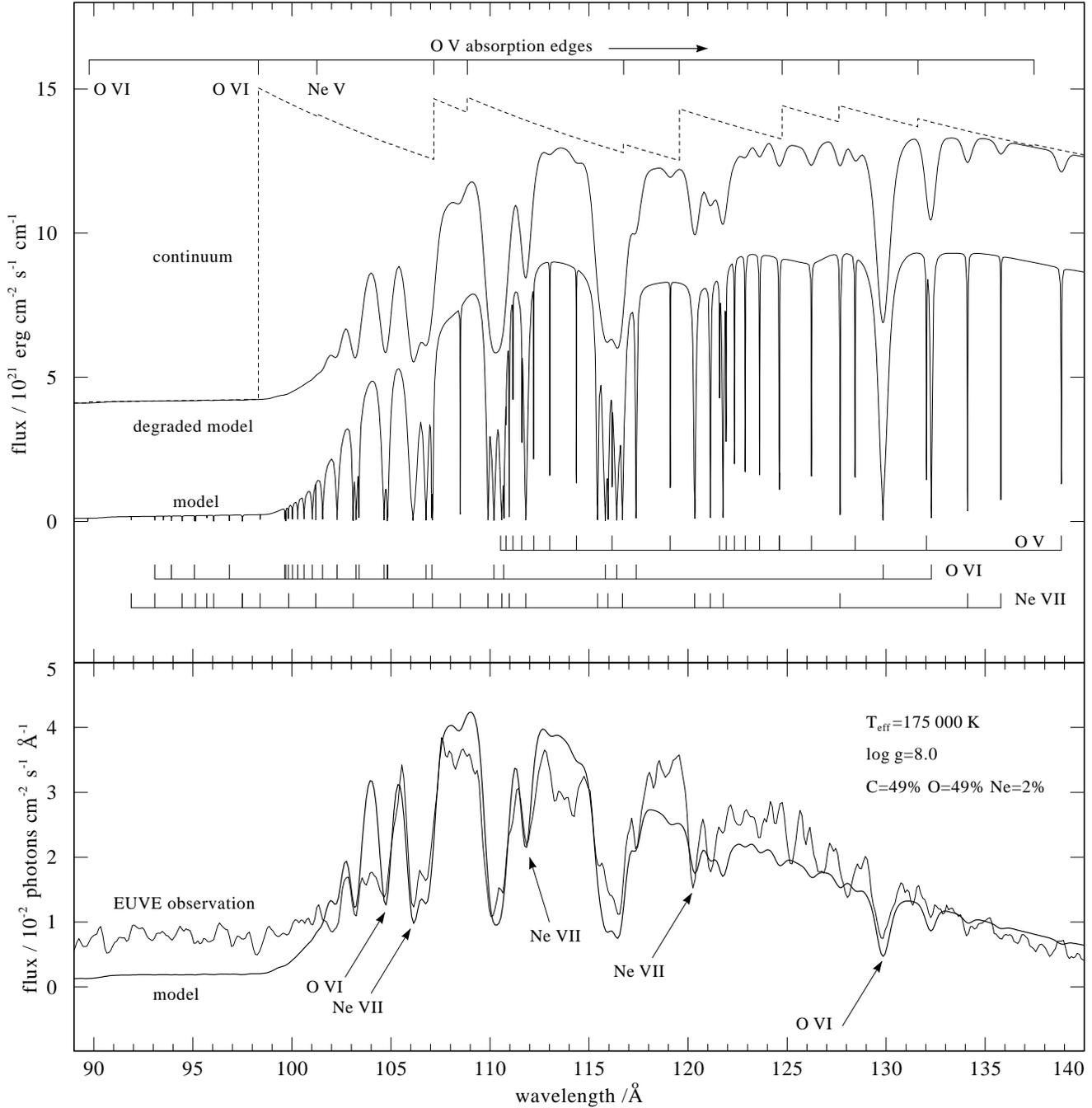}}
  \caption[]
{
Top panel: Emergent flux of a model with \Teff=175\,000\,K, \logg=8. Shifted
upward is the same model spectrum but degraded with a 0.5\AA\ FWHM Gaussian.
Overplotted (dashed line) is the continuum flux computed without occupation
probabilities in order to display the position of the absorption edges. They
are smeared out by atomic level dissolution and broad absorption lines. The
spectrum is dominated by strong and broad \ion{O}{VI} lines and less prominent
\ion{O}{V} and \ion{Ne}{VII} lines. Bottom panel: Comparison of the observed
EUVE spectrum with this model (attenuated by an interstellar column density
$N({\rm \ion{H}{I}})=5.1\cdot 10^{19}$cm$^{-2}$ with ${\rm \ion{He}{II}}/{\rm
\ion{H}{I}} = 0.068$ and ${\rm \ion{He}{I}}/{\rm \ion{H}{I}} = 0.052$)
}
\label{euve}
\end{figure*}

\section{Observations and data reduction}

The EUVE spectrum of \hh\ was retrieved from the public archive. The exposure
lasted 38\,000 seconds, starting on Dec.\,5, 1993. It was reduced (including
flux calibration and subtraction of higher order contributions) with the
standard procedures of the IRAF/EUV software package (Version 1.6.2). \hh\ is
only detected in the EUVE short wavelength spectrometer because of its
relatively high interstellar column density. The observed spectrum covers the
range 75--150\AA\ at a resolution of 0.5\AA.

A useful complement are X-ray ROSAT PSPC data, which were retrieved from
the public archive. The observation lasted 4882 seconds, beginning on July 18,
1990. A pulse height distribution was extracted by using the MIDAS-EXSAS
software. Source flux is detected in the range 30--120\AA\ (100--420\,eV).

We have performed optical spectroscopy of \hh\ at the 10m Keck~I telescope on
July 20, 1998, under excellent weather conditions. We have taken two spectra,
which were later co-added, with a total of 6000 seconds exposure time, using
the high resolution echelle spectrograph (HIRES) and the blue cross disperser.
For details on data reduction see Zuckerman \& Reid (1998). The spectrum covers
the full wavelength region between 3600\AA\ and 5100\AA\ at a resolution of
35\,000.

\section{Model atmospheres and spectral fitting}

We have computed a grid of non-LTE model atmospheres with different input
parameters (\Teff, \logg, element abundances). The models include the most
abundant elements, C and O, as well as Ne and He self-consistently and they are
plane-parallel and in radiative and hydrostatic equilibrium. The computer code
is based on the Accelerated Lambda Iteration method and is described in detail
by Werner \& Dreizler (1999).

Let us present the atomic input data at some length, because this is the first
detailed analysis of an EUV spectrum of a PG\,1159 star with data of this good
quality. We note in passing that similar models were applied only in the case
of one other PG\,1159 star (PG\,1520+525; Werner \etal 1996a) from which a
useful spectrum could be taken by EUVE. However, the S/N was much worse in that
case.

\subsection{Potential spectral lines and absorption edges in the EUVE spectrum
of \hh}

No features of \ion{He}{II} and \ion{C}{IV} are located in the 75--150\AA\
spectral range, however, \ion{C}{IV} absorption edges at longer wavelengths are
important because they contribute strongly to the background opacity in the
EUVE region. Although \ion{C}{V} is the dominant ionization stage of carbon we
do not expect to see any of its spectral lines because temperatures are too low
to populate excited \ion{C}{V} levels. On the other hand, we know from previous
studies that the \ion{C}{V} ground state edge (31.6\AA) almost completely
blocks the flux (Werner \etal 1996b) in a PG\,1159 stellar atmosphere. This has
no observable influence for the EUVE range, but is important in order to
interpret correctly the ROSAT PSPC data.

The strongest opacity source is oxygen. \ion{O}{VI} is most important: In the
spectral range in question we find two absorption line series arising from the
ground state and the first excited state. The respective absorption edges are
located at 89.8\AA\ and 98.3\AA\ (see Fig.\,\ref{euve}, top panel), the latter
one causes the flux to drop at $\lambda<100$\AA. Absorption edges of \ion{O}{V}
are weaker (because \ion{O}{VI} is more populated), but they are more numerous
so that their combined opacity is important for the overall flux distribution
(Fig.\,\ref{euve}). These edges arise from the six lowest \ion{O}{V} levels,
where we have accounted for the possibility that from any 2s2p configuration
either the 2p or 2s electron can be ionized, leaving behind an \ion{O}{VI} ion
in the ground state or first excited state, respectively; therefore two edges
appear at different locations in the spectrum. Spectral lines of \ion{O}{V} are
less important because of the weak population of this ion and because the line
profiles are intrinsically less broad than those of the \ion{O}{VI} lines
(quadratic vs. linear Stark effect, see below).

Many lines from \ion{Ne}{VII} are located in the 80-140\AA\ range, as well as
numerous absorption edges from \ion{Ne}{IV} to \ion{Ne}{VII}. The edges turned
out to be very weak as compared to oxygen, which is of course a consequence of
the lower Ne abundance.

Primary source for the used level energies is Bashkin \& Stoner (1975). Carbon
and oxygen oscillator strengths for spectral lines as well as bound-free
cross-sections for photon and electron collisional ionization were largely
provided by K.\ Butler (priv.\,comm.). Oscillator strengths for \ion{Ne}{VII}
lines were retrieved from the Opacity Project data base (Seaton \etal 1994), and
bound-free cross-sections were computed hydrogen-like (the Ne model atom is
essentially identical to that of Werner \& Rauch 1994). Alternatively we have
also used Opacity Project photon ionization cross-sections for all species, but
the consequences are not relevant for our present study.

Finally we note that helium, for which an upper limit of 1\% by mass was
derived, does not at all affect the EUV spectrum.

\subsection{Line broadening and pressure ionization}

Line broadening is a problem which can be tackled only approximately at best,
particularly in the most important case, the \ion{O}{VI} lines. \ion{O}{VI} is
a hydrogenic ion. Its energy levels with equal principal quantum number are
closely spaced but not degenerate. As a consequence, line broadening ranges
between the linear and quadratic Stark regimes. We faced the same problem for
optical lines in previous analyses and we refer to Werner \etal (1991) for
details on the approximative approach which we apply here, too. For \ion{O}{V}
and \ion{Ne}{VII} lines we assume quadratic Stark broadening, which is
reasonable except for the highest members of the \ion{O}{V} line series in our
synthetic spectra, because linear Stark effects could become important here,
too.

Another problem is posed by the numerical treatment of the bound-free photon
cross-sections. It is clear from the outset that absorption edges in a stellar
spectrum are not really sharp. Pressure effects lower the atomic ionization
potential and tend to smear them out (see Fig.\,\ref{euve}, top panel). We
account for this effect by an occupation probability formalism (Hummer \&
Mihalas 1988) which was generalized to NLTE conditions by Hubeny \etal (1994).
We use essentially their numerical treatment, but two details demand special
attention in our case. First, the perturber particles, which impose an electric
microfield at the location of the radiating atom, are not protons (as it is the
case in atmospheres with solar composition), but highly charged particles
(mostly \ion{C}{V} and \ion{O}{VII}). Consequently the critical field strength
$\beta_{\rm c}$ as given in Hubeny \etal (1994; their eq.\,A.2) needs to be
scaled by C$^{1/3}$, where C is the ratio of the total ion density to the
electron density, $n_{\rm Ion}/n_{\rm e}$. And second, the plasma correlation
parameter $a$, which is defined as the ratio of the mean distance of ions to
the Debye length, needs to be scaled, so that we have instead of their eq.\,A.4:
\[
a =
 0.09 \frac{n_{\rm e}^{1/6}}{\sqrt{T}} {\left(1 + \sum_i Z_{\rm i}^2 N_{\rm i}/n_{\rm e} \right)^{1/2}
\left( \sum_i N_{\rm i}/n_{\rm e}  \right)^{-1/3} }
\]
where $N_{\rm i}$ and $Z_{\rm i}$ are the number density and charge of the i-th
ion, respectively, and $n_{\rm e}$ and $T$ are the electron density and
temperature. This procedure is formulated strictly for hydrogenic ions only
and, hence, applicable for \ion{O}{VI}, but we also treat the \ion{O}{V} edges
in this way, because we lack any better approach. But since the \ion{O}{V}
edges are weak, this uncertainty is not important for our analysis.

To summarize, the most severe uncertainty in the synthetic spectrum
calculations is the lack of a satisfactory treatment of \ion{O}{VI} line
broadening, because strong \ion{O}{VI} line merging has a significant effect on
the overall shape of the EUV spectrum. Theoretical line broadening data exist
for only few of the \ion{O}{VI} lines relevant here, but these do not account
for linear Stark effects (Dimitrijevic \& Sahal-Brechot 1992).

\subsection{Iron group elements}

EUVE spectra of hot hydrogen-rich white dwarfs are completely dominated by iron
and nickel lines and bound-free continua (e.g.\ Wolff \etal 1998), hence, it
was felt mandatory to check the relevance of these opacities in the case of
\hh. Model calculations show that effects of the iron group elements (with a
solar abundance fraction) are not detectable at the resolution and S/N level of
the present EUVE spectrum. Continuous opacities of the Fe group are negligibly
small compared to the dominant species. On the other hand, our model predicts
that a large number of lines should be detectable in future observations with
the Chandra X-ray observatory, which will be able to provide spectra
with higher resolution and better S/N.

\subsection{Treatment of interstellar absorption in synthetic spectra}

EUV and soft X-ray spectra are very sensitive to absorption from interstellar
hydrogen and helium. For the EUVE spectrum this attenuation was calculated
according to the model of Rumph \etal (1994). The interstellar column density
of neutral hydrogen ($N({\rm \ion{H}{I}})$) enters as a free parameter. It was
chosen so that the observed flux could be reproduced at $\lambda \ga 130$\,\AA.
The column densities of \ion{He}{I} and \ion{He}{II} were fixed relative to
hydrogen using the mean values of ${\rm \ion{He}{I}}/{\rm \ion{H}{I}} = 0.068$
and ${\rm \ion{He}{II}}/{\rm \ion{H}{I}} = 0.052$ from Wolff \etal (1999).

For the ROSAT PSPC pulse height distribution we have also used fixed relative
abundances for helium (see Jordan \etal (1994) for details of the analysis).
Since different spectral regions are used to determine the interstellar
absorption the $N({\rm \ion{H}{I}})$ values from the ROSAT and EUVE analyses
differ.

\section{Results and discussion}

In the top panel of Fig.\,\ref{euve} we display a model spectrum for \hh. The
numerous O and Ne lines are identified and close inspection shows that the
\ion{O}{VI} lines are strongest. Line blending with \ion{O}{V} and
\ion{Ne}{VII} occurs at many locations and the degraded spectrum (shifted
upward in the Figure) which simulates the EUVE resolution of 0.5\AA\
demonstrates that disentangling of single lines becomes difficult. However, it
is possible to identify a few isolated lines or line cores of \ion{O}{VI} and
\ion{Ne}{VII} which are indicated by arrows in the lower panel of
Fig.\,\ref{euve}. The roll-over of the flux towards shorter wavelengths is
caused by the strong absorption edge of the first excited level of \ion{O}{VI}
at 98\AA, which can be seen by the dashed line in the top panel of
Fig.\,\ref{euve} which represents the continuum flux of the model. As already
mentioned, the sharp absorption edge is strongly smoothed out by pressure
effects as well as the converging absorption line series. For the same reason
all other absorption edges which are indicated in Fig.\,\ref{euve} cannot be
recognized in the final model spectrum.

\subsection{Effective temperature}

The best fitting model was found by taking models with different \Teff\ and
keeping fixed \logg=8 and the abundance ratio C/O=1. The latter two parameters,
when varied within the limits of the optical line analysis, affect the model
spectrum much less than \Teff. The model flux is normalized to the observed
visual magnitude (V=16.24, Nousek \etal 1986) and the interstellar column
density, which is responsible for the roll-off at the long wavelength end of
the spectrum, is treated as a free parameter. The neon abundance can be
estimated from the \ion{Ne}{VII} line near 106\AA\ which is observed to have a
deeper core than the blending \ion{O}{VI} lines near 107\AA.

The best fit is displayed in the bottom panel of Fig.\,\ref{euve}, with model
parameters as indicated. The overall fit is satisfactory. The strongest
\ion{O}{VI} and \ion{Ne}{VII} line features can be identified and the model can
fit them. The absolute flux level at the flux maximum and at longer wavelengths is
matched by the model with \Teff=175\,000\,K. Models with \Teff\ less than
170\,000\,K can be clearly ruled out, because the \ion{O}{V} absorption edges
become too strong. On the other hand, \Teff\ higher than 180\,000\,K causes a
strong flux excess in the region of the observed flux maximum.

\begin{figure}
  \resizebox{\hsize}{!}{\includegraphics{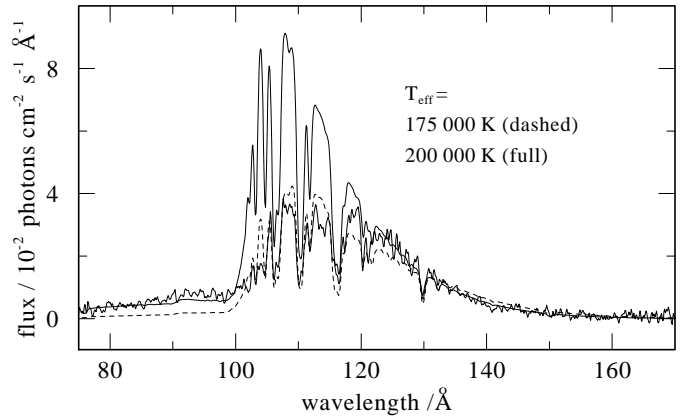}}
  \caption[]
{
EUVE spectrum as compared to two models with different \Teff: 175\,000\,K
(dashed line) and 200\,000\,K (full line). The hot model fits the flux at
$\lambda<100$\AA, but overestimates the maximum flux by a factor of two.
Interstellar columns are  $N({\rm \ion{H}{I}})=5.1\cdot 10^{19}$cm$^{-2}$ and
$6.0\cdot 10^{19}$cm$^{-2}$, respectively. Other model parameters as in
Fig.\,\ref{euve}
}
\label{euve_hot}
\end{figure}
\begin{figure}
  \resizebox{0.83\hsize}{!}{\includegraphics{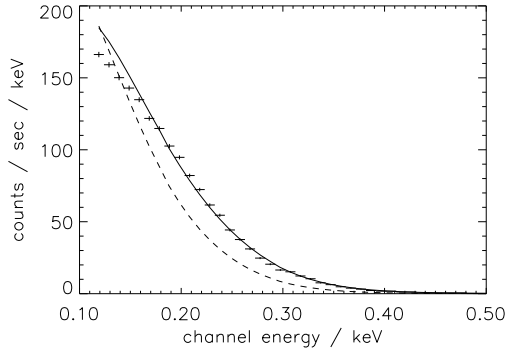}}
  \caption[]
{
The same models which were fitted to the EUVE spectrum in Fig.\,\ref{euve_hot}
are here compared to the ROSAT PSPC pulse height distribution. The 175\,000\,K
model (dashed line) cannot fit the data but the 200\,000\,K model (full line)
does, except for the softest channels (see text for discussion). Interstellar
columns are $N({\rm \ion{H}{I}}) = 5.75\cdot 10^{19}$cm$^{-2}$ and $9.5\cdot
10^{19}$cm$^{-2}$, respectively
}
\label{rosat}
\end{figure}

However, in one respect this model fails to fit the observation. Its flux is
distinctly too low at $\lambda<100$\AA. This could in principle be improved
with a hotter model (200\,000\,K, along with a higher column density to fit the
spectrum at long wavelengths), but the flux exceeds the maximum by a factor of
two in the range 100--120\AA\ (Fig.\,\ref{euve_hot}). This is the range where
strong \ion{O}{VI} line merging occurs and it is conceivable that, in case that
we underestimate the \ion{O}{VI} line broadening, stronger line wings can
effectively block more flux in this region. Alternatively, yet unidentified
opacity sources of other light metals may be responsible (e.g.\ Mg might be as
abundant as Ne, and many \ion{Mg}{V} and \ion{Mg}{VI} lines are located in the
EUV), but any progress must await data with better spectral resolution.

Another argument that speaks for \Teff\ slightly higher than 175\,000\,K is the
strength of the optical \ion{Ne}{VII}~3644\AA\ line (the very strong EUV lines
of Ne are much less temperature sensitive). A better fit is obtained at
200\,000\,K, although still, the equivalent width is too weak
(Fig.\,\ref{keck}). This can be improved by reducing \logg\ to 7.5 and/or
increasing the Ne abundance (see below).

As one can expect from the EUVE flux below 100\AA, the ROSAT PSPC pulse height
distribution (120--30\AA) cannot be matched by the ``cool'' 175\,000\,K model
(Fig.\,\ref{rosat}). A model with \Teff=200\,000\,K is needed to obtain a
satisfactory fit (where we accept a model flux excess at the softest energies,
which might be caused by the underestimated flux blocking in the 100-120\AA\
range as mentioned above).

To summarize,  we estimate \Teff=170\,000--200\,000\,K from the EUV and X-ray
data, which compares reasonably well with the range obtained previously from
the optical spectra (\Teff=150\,000--190\,000\,K). This indicates that \hh\
approaches or even exceeds the temperature of the hottest known PG\,1159 star
(\Teff=180\,000$\pm$20\,000\,K; RX\,J0122.9$-$7521; Werner \etal 1996b).

\subsection{Neon abundance}

\begin{figure}
  \resizebox{\hsize}{!}{\includegraphics{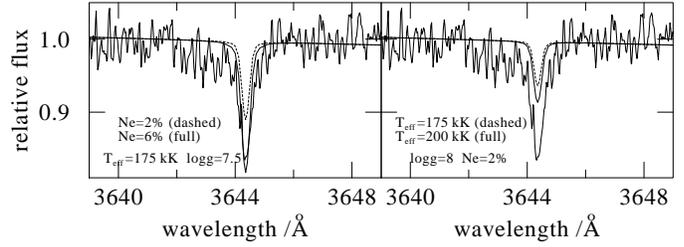}}
  \caption[]
{
The Keck spectrum shows the \ion{Ne}{VII}~3644\AA\ line and
its strength confirms the high neon abundance deduced from the EUV lines. Left
panel: Profiles from the two models fitted to EUVE and ROSAT data in
Figs.\,\ref{euve_hot} and \ref{rosat}. With Ne=2\% (20 times solar value) the
profiles are still too weak, however, they are also gravity dependent, as can
be seen in the right panel (\logg\ decreased from 8 to 7.5). It suggests that
the Ne abundance might be even higher than 2\%
}
\label{keck}
\end{figure}

The EUV spectrum indicates a high neon abundance (2\%), which is a factor of 20
higher than the solar value. We estimate a possible error of a factor of three,
which is caused by the S/N of the data and the fact that the strong
\ion{Ne}{VII} lines are saturated. Our first detection of the
\ion{Ne}{VII}~3644\AA\ line in the Keck spectrum of \hh\ (a previous attempt
with the 3.5m Calar Alto telescope failed) indicates, that the actual abundance
might be even higher. Fig.\,\ref{keck} shows several model fits to this line.
Even under the assumption of a lower gravity (\logg=7.5), which represents the
lower limit of the optical analysis and which results in a stronger line, the
relatively strong observed profile suggests an abundance of the order 5\%. We
successfully detected this line earlier in three other PG\,1159 stars (Werner
\& Rauch 1994) and, as discussed in detail in that paper, the derived Ne
abundance (2\%) can be expected in matter processed by 3$\alpha$ burning. This
high abundance can be rated as an indication that the stars have lost their H-
and He-rich envelopes, with \hh\ being the most extreme case.

We have already speculated that the He-deficiency in \hh\ points at the
possibility that the star has burned carbon in previous evolutionary stages
(Werner 1991). Hence, it might have been one of the ``heavyweight''
intermediate-mass stars (8\,M$_{\sun}\lappr$ M $\lappr$ 10\,M$_{\sun}$) which
form white dwarfs with electron-degenerate O-Ne-Mg cores resulting from carbon
burning. The extraordinarily high Ne abundance found in the present study and
the unusually high mass of \hh\ corroborate this speculation. In any case, the
Mg and Na abundances can be as high as the Ne abundance (Iben \etal 1997), so that
better EUV spectra, which can be provided by the future Chandra mission, should
be used for further studies in this direction with a possible identification of
other metal lines.

\begin{acknowledgements}
Keck data were taken in collaborative observations with I.N.\,Reid. A travel
grant from the DFG (We\,1312/21-1) to the Keck Observatory is gratefully
acknowledged. ROSAT data analysis in T\"ubingen and Kiel is supported by the
DLR under grants 50\,OR\,97055 and 50\,OR\,96173, respectively. Correspondence
with I.\,Hubeny on occupation probabilities was helpful.
\end{acknowledgements}


\begin{thebibliography}{}

\bibitem{Barstow95} Barstow M.A., Holberg J.B., Koester D., Nousek J.A., Werner K. 1995,
        in White Dwarfs, eds.\ D. Koester and K. Werner, Lecture Notes in
        Physics 443, Springer, Berlin,
        p. 302

\bibitem{BS75} Bashkin S., Stoner J.O., Jr. 1975, Atomic Energy Levels \&
	Grotrian Diagrams, Vol.\ 1, Amsterdam, North Holland

\bibitem{DS92} Dimitrijevic M.S., Sahal-Brechot S. 1992, A\&AS 93, 359

\bibitem{Hub94} Hubeny I., Hummer D.G., Lanz T. 1994, A\&A 282, 157

\bibitem{HuMi88} Hummer D.G., Mihalas D. 1988, ApJ 331, 794

\bibitem{Iben97} Iben I. Jr., Ritossa C., Garcia-Berro E. 1997, ApJ 489, 772

\bibitem{Jordan94} Jordan S., Wolff B., Koester D., Napiwotzki R. 1994,
        A\&A 290, 834

\bibitem{KrW98} Kruk J.W., Werner K. 1998, ApJ  502, 858

\bibitem{Nousek96} Nousek J.A., Shipman H.L., Holberg J.B., Liebert J., Pravdo S.H.,
	White N.E., Giommi P. 1986, ApJ 309, 230

\bibitem{Rumph94} Rumph T., Bowyer S., Vennes S. 1994, AJ 107, 2108

\bibitem{Sea94} Seaton M.J., Yan Y., Mihalas D., Pradhan A.K. 1994, MNRAS 266, 805

\bibitem{KW91} Werner K. 1991, A\&A 251, 147

\bibitem{WR94} Werner K., Rauch T. 1994, A\&A 284, L5

\bibitem{WD99} Werner K., Dreizler S. 1999, in Computational Astrophysics, eds.\ H. Riffert and K.
        Werner, Journal of Computational and Applied Mathematics, Elsevier,
        in press

\bibitem{WHH91} Werner K., Heber U., Hunger K. 1991, A\&A 244, 437

\bibitem{Wetal96a} Werner K., Dreizler S., Heber U., Rauch T. 1996a,
        in Astrophysics in the Extreme Ultraviolet, IAU Colloquium 152,
        eds.\ S. Bowyer and R.F. Malina, Kluwer, p. 229

\bibitem{Wetal96b} Werner K., Wolff B., Pakull M., Cowley A.P., Schmidtke P.C.,
        Hutchings J.B., Crampton D. 1996b, in Supersoft X-ray Sources, ed.\ J. Greiner, Lecture 
	Notes in Physics 472, Springer, Berlin, p. 131

\bibitem{Wolff98} Wolff B., Koester D., Dreizler S., Haas S. 1998, A\&A 329, 1045

\bibitem{Wolff99} Wolff B., Koester D., Lallement R. 1999, A\&A in press

\bibitem{ZR98} Zuckerman B., Reid I.N. 1998, ApJ 505, L143

\end{thebibliography}
\end{document}